\input harvmac
\input epsf

\def\half{{1\over 2}}

\def\wn{\omega_{n-1}}
\def\({\left(}
\def\){\right)}

\Title{}{\vbox{\centerline{One Conjecture and Two Observations on
de Sitter Space}}}

\centerline{Qing-Guo Huang$^1$, Ke Ke$^2$ and Miao Li$^{1,2}$}

\medskip
\centerline{\it $^1$ Interdisciplinary Center of Theoretical
Studies} \centerline{\it Academia Sinica, Beijing 100080, China}
\medskip
\centerline{\it and}
\medskip
\centerline{\it $^2$ Institute of Theoretical Physics}
\centerline{\it Academia Sinica, P. O. Box 2735} \centerline{\it
Beijing 100080}

\bigskip

\centerline{\tt huangqg@itp.ac.cn} \centerline{ \tt kek@itp.ac.cn}
\centerline{\tt mli@itp.ac.cn}

\bigskip

We propose that the state represented by the Nariai black hole
inside de Sitter space is the ground state of the de Sitter gravity,
while the pure de Sitter space is the maximal energy state. With
this point of view, we investigate thermodynamics of de Sitter
space, we find that if there is a dual field theory, this theory can
not be a CFT in a fixed dimension. Near the Nariai limit, we
conjecture that the dual theory is effectively an $1+1$ CFT living
on the radial segment connecting the cosmic horizon and the black
hole horizon. If we go beyond the de Sitter limit, the ``imaginary"
high temperature phase can be described by a CFT with one dimension
lower than the spacetime dimension. Below the de Sitter limit, we
are approaching a phase similar to the Hagedorn phase in $2+1$
dimensions, the latter is also a maximal energy phase if we hold the
volume fixed.

\Date{May, 2005}


\nref\ew{E. Witten, hep-th/0106109. }

\nref\dls{T. Banks, hep-th/0007146; M. Li, hep-th/0106184, JHEP 0204 (2002) 005;
T. Banks, astro-ph/0305037;
M. K. Parikh and E. Verlinde, hep-th/0403140; hep-th/0410227, JHEP 0501 (2005) 054.}

\nref\gh{G. W. Gibbons and S. W. Hawking, Phys. Rev. D 15 (1977)
2738. }

\nref\mo{Y. Myung, Mod.Phys.Lett. A16 (2001) 2353; Y. Myung,
Phys.Lett. B579 (2004) 205-210; S. Nojiri and S. Odintsov,
Phys.Lett. B523 (2001) 165-170; M. I. Park, Phys. Lett. B440,275
(1998); M. I. Park, Nucl. Phys. B544, 377 (1999). }

\nref\bbm{V. Balasubramanian, Jan de Boer and D. Minic, Phys. Rev.
D 65 (2002) 123508, hep-th/0110108. }

\nref\rb{R. Bousso, JHEP 0011:038, 2000  hep-th/0010252. }

\nref\bh{R. Bousso and S. W. Hawking, Phys. Rev. D 57 (1998) 2436,
hep-th/9709224. }

\nref\ssv{M. Spradlin, A. Strominger and A. Volovich,
hep-th/0110007. }

\nref\as{A. Strominger, JHEP 0110 (2001) 034, hep-th/0106113; A.
Strominger, JHEP 0111 (2001) 049. }

The quantum theory of gravity in de Sitter space is a longstanding
problem in modern physics. It may even be the most important problem for
the quantum gravity theorist, if the fate of our universe is indeed a
de Sitter space. There are two challenging issues in building a quantum theory of
de Sitter gravity. If we are to base the theory on the foundation of quantum
field theory or string theory, it appears impossible to define
measurable observables in an asymptotically de Sitter space \ew. On the other hand,
the holographic principle seems to imply that the number of states in the de Sitter
gravity is finite, its logarithmic is given by the Gibbons-Hawking entropy,
again, a field theory or string theory has no such features. Currently there are
two viewpoints regarding these problems. One viewpoint simply presumes that our universe
is not asymptotically de Sitter, therefore there do not exist these problems at
all, another view is the opposite, and some extreme form even postulates that
a small and positive cosmological constant is closely related to supersymmetry
breaking. We shall in this note take the second viewpoint. There exists considerable
amount of work based on the second viewpoint, for a partial list,
see \dls\ and \refs{\ew, \bbm,\as}.

An important step in approaching the quantum gravity theory in de
Sitter space is to work out thermodynamics of de Sitter space. The
difficulty here is that there is only one de Sitter space once the
cosmological constant of the theory is chosen, thus the usual
procedure of calculating thermodynamical quantities is not
applicable. However, a family of solutions labelled by the mass of
black hole exists, by varying the mass parameter we can vary the
cosmic horizon \refs{\gh, \mo}. We shall accept the N-bound here,
thus the pure de Sitter state is the maximal entropy state with the
maximal temperature. What is implicit in the literature but not
explicitly stated is that the Nariai black hole is the lowest energy
state, namely the ground state. Indeed for the Nariai black hole,
the cosmic horizon and the black hole horizon coincide and the
Gibbons-Hawking temperature vanishes. We will regard the entropy of
the Nariai black hole as the ground state degeneracy, and consider
other states with a larger entropy as excited states, in particular,
the pure de Sitter state is the limiting state with the largest
entropy. Thus, in our thermodynamic considerations, we will subtract
the ground state energy as well as the ground state entropy.
Subtracting the ground state entropy is an unusual procedure, by
doing this we implicitly assume that the entropy of excited states
is additive, one part is from the ground state degeneracy, another
is due to excited degrees of freedom. We suggest that the subtracted
energy and entropy encode the relevant information about the de
Sitter gravity.

In this short note, we calculate the subtracted energy and entropy
in four dimensional spacetime. Taking three limits, we make one
conjecture in near the Nariai limit, one observation in near the
pure de Sitter space limit and another observation in the imaginary
high temperature limit. We also generalize our considerations in
four dimensions to the cases in higher and $2+1$ dimensions.

Let us start with the metric of Schwarzschild-de Sitter solution
in four dimensional spacetime \eqn\dess{ds^2=- f(r) dt^2 +
f^{-1}(r)dr^2 + r^2 d \Omega_2^2, } with \eqn\fr{f(r)=1-{r_0 \over
r}-{r^2 \over L^2}, }
where $r_0=2Gm$ and $G$ is the Newton
constant. Here $m$ is an integration constant and
$L=\sqrt{3/\Lambda}$ is the size of the pure de Sitter space with a
positive cosmological constant $\Lambda$. When $m>0$, a horizon of
black hole appears. Raising the parameter $m$, the size of the
black hole horizon will increase, while the size of the cosmic
horizon will decrease. When $m=m_N={1 \over 3 \sqrt{3}} {L \over
G}$, or $r_0={2 \over 3 \sqrt{3}}L$, the black hole horizon will
be coincident with the cosmic horizon with size
$r_{c}=r_{BH}=r_N=L/\sqrt{3}$. This is the Nariai black hole.

The temperature of the cosmic horizon of the solution \dess\ is
given by \eqn\tempc{T_c={1 \over 4 \pi r_c} \left(3 {r_c^2 \over
L^2} - 1 \right), } and the relation between parameter $m$ and the
size of the cosmic horizon is \eqn\mc{m={r_c \over 2G}
\left(1-{r_c^2 \over L^2} \right). } The cosmic entropy is
determined by the the cosmic horizon area \eqn\entc{S_c={\pi r_c^2
\over G}. } In \bbm, Balasubramanian et al. use the surface
counter term method to find the energy of four dimensional
Schwarzschild-de Sitter spacetime and the result is $E=-m$. This
result is in accordance with the first law of thermodynamics
for the cosmic horizon which is given by \eqn\fl{d(-m)=T_c d S_c.
} If we choose a positive temperature, the physical energy in de
Sitter space should be $-m$, the same as the result in \bbm.

Bousso proposed that in any asymptotically de Sitter spacetime,
the total observable entropy is bounded by the pure de Sitter
entropy \rb. The authors of \bbm\ also put forward a conjecture
stating that any asymptotically de Sitter space whose mass exceeds
that of the pure de Sitter space contains a cosmological
singularity. These authors also proposed that the state
corresponding to the Nariai black hole has the lowest energy and
minimum entropy. We see from eq.\tempc\ that the temperature of
the cosmic horizon is zero in the Nariai limit. Based on this
fact, we suggest that the Nariai solution represents the ground
state of de Sitter spacetime. A state with an energy greater
than the Nariai mass is an excited state in de Sitter space.

In general, a thermal system trends to a state with largest entropy
if the energy is fixed, or, trends to a state with lowest energy if
the entropy is fixed. For the Schwarzschild-de Sitter system, the
reliable function of state is free energy \eqn\fr{F=E-TS=-m-T_c
S_c=-{r_c \over 4 G} \left(1+{r_c^2 \over L^2} \right), } here we
use eq. \tempc, \mc\ and \entc. This system trends to be a state
with lowest free energy, namely the pure de Sitter state, which is
consistent with \bh.

We now subtract from the energy of a Schwarzschild-de Sitter
spacetime the ground state energy. The deviations of the physical
mass and the cosmic entropy are given by \eqn\dm{\bigtriangleup
M=M-M_N=m_N-m={1 \over 3 \sqrt{3} } {L \over G} + {r_c \over 2G}
\left({r_c^2 \over L^2}-1 \right) } and \eqn\dent{\bigtriangleup S_c
=S-S_N= {\pi \over G} \left(r_c^2 - {L^2 \over 3} \right), } where
$S_N=\pi r_N^2/G$ is the entropy of the cosmic horizon of the Nariai
black hole and here we use eqs.\mc\ and \entc. We suggest that these
subtracted thermodynamical quantities of the cosmic horizon encode
relevant information about the dual theory of the de Sitter gravity.
The first law of thermodynamics \fl\  can be re-expressed as
\eqn\therm{d (\bigtriangleup M) = T_c d (\bigtriangleup S_c). } The
size of the cosmic horizon can be regarded as a function of the
temperature through  eq.\tempc \eqn\rc{r_c=r_N \left[{T_c \over T_s}
+ \sqrt{\left({T_c \over T_s} \right)^2+1} \right], } where
$T_s=1/(2 \pi r_N)=\sqrt{3}/(2 \pi L)$. Then the subtracted energy
and the entropy are related to the temperature via
\eqn\dmt{\bigtriangleup M={2 L \over 3 \sqrt{3} G} \left({T_c \over
T_s} \right)^2 \left[{T_c \over T_s} + {\left({T_c \over T_s}
\right)^2 + \half + \sqrt{\left({T_c \over T_s} \right)^2 +1} \over
\sqrt{\left({T_c \over T_s} \right)^2 +1} +1 } \right] } and
\eqn\dentr{\bigtriangleup S_c = {2 \pi L^2 \over 3 G} {1 \over
\sqrt{\left({T_s \over T_c} \right)^2 +1} -1}. } Now, both $\Delta
M$ and $\Delta S$ are not simple scaling functions of the
temperature. we deduce that the quantum theory dual to  gravity in
de Sitter space can not be described by a single CFT theory. Of
course, there must be states of the same energy other than those
represented by Schwarzschild-de Sitter solutions, it appears that
our conclusion is too strong. However, we believe that the states of
Schwarzschild-de Sitter spacetime for a given energy dominates the
number of states of the same energy. Note that in the dS/CFT
correspondence conjecture \as, the CFT lives in Euclidean space, our
conclusion does not apply to that conjecture. We shall study thermal
properties of three limiting cases to make one conjecture and two
observations in the following.

\noindent $\bullet$ Near the Nariai limit - one conjecture

First, we study the system near the Nariai hole whose temperature is
much smaller than $T_s$. The temperature of the cosmic horizon is
almost the same as that of the black hole horizon, more accurately,
the difference between these two temperatures is proportional to
$T_c^2$. It says that this system is almost in thermal equilibrium.
One might think that the region between the two horizons will
collapse to a sphere in the Nariai limit, this turns out not the
case. The proper distance between the cosmic horizon and the black
hole horizon is \eqn\pd{\ell = {L \over \sqrt{3}}
\int_{r_{BH}}^{r_c} [(r_c - r)(r - r_{BH})]^{-1/2} dr=\pi r_N, }
here we take $r_{BH} \rightarrow r_c$ in the last step. The
subtracted energy and the entropy are given by
\eqn\nndm{\bigtriangleup M = {2 \pi^2 \over 3 \sqrt{3}} {L^3 \over
G} T_c^2} and \eqn\nndentt{\bigtriangleup S_c = {4 \pi^2 \over 3
\sqrt{3}} {L^3 \over G} T_c. } These relations can be interpreted as
arising in a conformal field theory (CFT) in $1+1$ dimensional
spacetime. We conjecture that in the near extremal (Nariai) limit,
the physics can be described by a 2d CFT. Now the question is, where
does this CFT live? It is interesting to notice that in the near
extremal limit, the the spatial part of the metric reduces to
$ds^2=d\rho^2+r_N^2 d\Omega_2^2$, where $\rho=2r_N \sin^{-1} \left(
\sqrt{r-r_- \over r_+-r_- }\right)$. Not only the topology, but also
the geometry factorizes into the form $I \times S^2$. It appears
that the only natural choice is the segment between the two
horizons. There is a universal relation between the energy and the
temperature in a $1+1$ dimensional CFT which reads $\bigtriangleup
M={\pi c \over 6} \ell T^2$, where $c$ is the central charge of the
CFT and $\pi R=\ell$, $\ell$ is the proper length of the spatial
dimension. Applying this relation to eq. \nndm, we read off the
central charge of this CFT to be $c={12 \over \pi} S_N$. To specify
the CFT, we also have to impose boundary conditions at the two ends
of the segment, but we do not have enough information to fix these
conditions. However, the thermal properties will not depend on the
choice of boundary conditions. Here we stress that Schwarzschild-de
Sitter system is not a conservative system and the energy is not
conserved. The dual CFT is living in the segment and the energy can
also flow outwards or inwards.

To sum up, our conjecture is that {\it near the Nariai limit, the
quantum gravity in dS space is dual to an $1+1$ CFT living on the
radial segment connecting the cosmic horizon and the black hole
horizon.}

\noindent $\bullet$ Near the pure de Sitter limit - one observation

According to the N-bound conjecture, the state with the highest
entropy of this system is the pure de Sitter case, and the
temperature is also the highest. Using eqs.\dmt\ and \dentr, we find
that near the pure de Sitter state the subtracted physical mass and
entropy are approximately \eqn\ndsm{\bigtriangleup M={L \over 3
\sqrt{3} G} + {L \over 2G }{T_c - T_{ds} \over T_{ds}} } and
\eqn\ndsentr{\bigtriangleup S_c = {2 \pi L^2 \over 3 G} + {\pi L^2
\over G }{T_c - T_{ds} \over T_{ds}}, } where $T_{ds}=1/(2 \pi L)$
is the temperature of the cosmic horizon in pure de Sitter space.

The pure de Sitter state is a limiting state, which reminds us of
the Hagedorn temperature in string theory. In the canonical ensemble
approach, the string soup can not be heated above the Hagedorn
temperature. Near this temperature, the string system tends to form
a single long string with the energy scales as \eqn\stre{E\sim
{1\over T_H-T},} where $T_H$ is the Hagedorn temperature. This
behavior is different from the de Sitter limit where the energy
approaches a constant. However, states in de Sitter space are always
confined within the cosmic horizon, thus from whatever point of
view, the system is one in a finite volume, while the volume
occupied by a Hagedorn string diverges. One can estimate the volume
occupied by the string as follows. A long string looks like a random
walk, so its size $R$ scales with its actual length as $R\sim
\sqrt{L}$, the energy of the string is proportional to the length
$L$ and the former scales with the oscillator level $N$ as
$\sqrt{N}$, thus the size of the string scales with the oscillator
level as $R\sim N^{1/4}$. At a temperature below $T_H$,
$\sqrt{N}\sim 1/(T_H-T)$. If the string lives in a $d$ dimensional
space, the volume occupied by the string behaves as $V=R^d\sim
1/(T_H-T)^{d/2}$. We see that when $d=2$, the volume has the same
divergence as the total energy. The entropy of the string is
proportional to $\sqrt{N}$ thus has the same divergence of the
energy. If we hold the volume fixed (thus there is only a segment of
string in this volume), both the energy and the entropy are also
fixed when $d=2$. The string with a fixed volume behaves just like
the de Sitter state where the temperature, the energy and the
entropy all have a finite limit. Moreover, we expect that the
corrections to the energy and the entropy both tend to zero linearly
in $T_H-T$.

In conclusion, our first observation is that {\it the de Sitter
system near the pure de Sitter phase behaves just like the Hagedorn
string in a fixed volume in $2+1$ dimensions.}

\noindent $\bullet$ Imaginary high temperature limit - another
observation

What is interesting about the de Sitter gravity is that
thermodynamics can be analytically extended to temperatures higher
than the pure de Sitter temperature. A higher temperature is
realized by the cosmic temperature of a spacetime when parameter $m$
becomes negative. In this spacetime, $r=0$ is a naked singularity.
It is convenient to replace $m$ by $-m$ to make $m$ positive, or
equivalently $r_0 \rightarrow -r_0$. Taking a limit with $r_0 \gg
L$, the size of the cosmic horizon is approximately given by
\eqn\mmrc{r_c\simeq (r_0 L^2)^{1/3}.} The temperature of the cosmic
horizon becomes \eqn\mmtc{T_c = {3 \over 4 \pi L} \left({r_0 \over
L} \right)^{1/3}, } and we have $T_c \gg T_s$. Eqs.\dmt\ and \dentr\
become \eqn\mmdm{\eqalign{\bigtriangleup M &= {32 \pi^3 L^4 \over 27
G} T_c^3\cr \bigtriangleup S_c &= {16 \pi^3 L^4 \over 9G } T_c^2. }}
These thermal relations are just the ones in a $2+1$ dimensional
CFT. Of course as we already said that there is a naked singularity
at $r=0$ not allowed by the cosmic censorship principle. So we can
get our second observation that {\it the imaginary high temperature
phase is described by a $2+1$ dimensional CFT. } When the
temperature drops below the de Sitter temperature, comparing \mmdm\
with \dentr, we find that the entropy of this $2+1$ dimensional CFT
is always less than that of the $dS_4$. This fact tells us that the
number of degrees of freedom in this $2+1$ CFT is not large enough
to describe the quantum theory of gravity in dS$_4$.

Next we will generalize our previous considerations to a higher
dimensional de Sitter space. Let the spacetime dimension be $n+1$.
The metric of the Schwarzschild-de Sitter solution takes the form
\eqn\nmt{ds^2=-f(r) dt^2 + f^{-1}(r) dr^2+r^2 d \Omega_{n-1}^2, }
where \eqn\nfr{f(r)=1-{r_0^{n-2} \over r^{n-2}} -{r^2 \over L^2},}
\eqn\rwn{r_0^{n-2}=\wn m, \quad \wn={16 \pi G \over (n-1)
Vol(S^{n-1})}, } $L=\sqrt{{n(n-1) \over 2 \Lambda}}$ is the size of
the pure dS$_{n+1}$ and $m$ is an integration constant. When $m
> 0$, there is a cosmic horizon and a black hole horizon. When $m$
increases, the size of the cosmic horizon decreases, but the size of
black hole increases. When $m$ reaches a critical value $m_N={2
\over n \wn} \left({n-2 \over n } L^2 \right)^{n-2\over 2}$,
corresponding to the Nariai solution, the cosmic horizon and the
black hole horizon share the same hypersurface with size
$r_c=r_{BH}=r_N=\sqrt{n-2 \over n} L$. Using metric \nmt, we can
express parameter $m$, the temperature and entropy corresponding to
the cosmic horizon as \eqn\mts{\eqalign{m&={r_c^{n-2} \over \wn }
\left(1-{r_c^2 \over L^2} \right), \cr T_c&={1 \over 4 \pi r_c}
\left[n{r_c^2 \over L^2}-(n-2) \right]={n \over 4 \pi L^2 r_c}
\left(r_c^2-r_N^2 \right), \cr S_c&={r_c^{n-1} Vol(S^{n-1}) \over 4
G}={4 \pi r_c^{n-1} \over (n-1) \wn }. }} The first law of
thermodynamics can be expressed as \eqn\nfl{d(-m)=T_c d S_c. } In
\bbm, the authors calculate the mass of the gravitational field in
$(4+1)$ dimensional spacetime and find $M_5={3 \pi L^2 \over 32 G
}-m$ by the counter term method. In higher dimensions, more counter
terms are needed. Since we only need to know the difference between
the energy of a general Schwarzschild-de Sitter spacetime and that
of the Nariai hole,  we can ignore the constant correction to the
mass. From eq.\mts, the subtracted mass and the entropy are given by
\eqn\ndms{\eqalign{\bigtriangleup M&={r_N^{n-2} \over \wn} \left[{2
\over n} + {n-2 \over n}\left({T_c \over T_s} + \sqrt{\left({T_c
\over T_s} \right)^2+1} \right)^n -\left({T_c \over T_s} +
\sqrt{\left({T_c \over T_s} \right)^2+1} \right)^{n-2} \right], \cr
\bigtriangleup S_c&= {4 \pi r_N^{n-1} \over (n-1) \wn}
\left[\left({T_c \over T_s} + \sqrt{\left({T_c \over T_s}
\right)^2+1} \right)^{n-1}-1 \right], }} where $T_s={n-2 \over 2 \pi
r_N}$.

Taking the near Nariai limit, $T_c \ll T_s$, the proper distance
between the cosmic and black hole horizons approaches $\ell=\pi
r_N/\sqrt{n-2}$. The subtracted mass and entropy in this limit read
\eqn\nms{\eqalign{\bigtriangleup M&= { n-1 \over \sqrt{n-2}} S_N
\ell T_c^2, \cr \bigtriangleup S_c&={2(n-1)\over \sqrt{n-2}} S_N
\ell T_c, }}
 where $S_N={4 \pi r_N^{n-1} \over
(n-1) \wn}$ is the entropy associated with the cosmic horizon in the
Nariai solution. These quantities describe a system excited over the
Nariai ground state. The metric also factorizes $ds^2=d\rho^2+r_N^2
d \Omega_{n-1}^2$, where $\rho={2 r_N \over \sqrt{n-2}}\sin^{-1}
\left(\sqrt{r-r_- \over r_+-r_-} \right)$. Again, eq.\nms\ tells us
that the excitation near the Nariai system can be described by a two
dimensional CFT  with the central charge $c={6(n-1) \over \pi
\sqrt{n-2}} S_N$. When $m$ is small and the spacetime is close to
the pure de Sitter space, the subtracted mass and the entropy are
approximated by \eqn\npds{\eqalign{\bigtriangleup M&={r_N^{n-2}
\over \wn} \left[{2 \over n} + {2 \over n-1} \left({n \over n-2}
\right)^{n-2 \over 2} {T_c -T_{ds} \over T_{ds}} \right], \cr
\bigtriangleup S_c&= {4 \pi r_N^{n-1} \over (n-1) \wn}
\left[\left({n \over n-2} \right)^{n-1 \over 2 }-1+ \left({n \over
n-2} \right)^{n-1 \over 2} {T_c-T_{ds} \over T_{ds}} \right], }}
where $T_{ds}=1/(2 \pi L )$ is the temperature of the cosmic horizon
in pure de Sitter space. The scaling behavior once again is similar
to the Hagedorn string in a fixed volume in 2+1 dimensions.

We also like to examine the unphysical high temperature phase  with
$r_0^{n-2} \rightarrow -r_0^{n-2}$ and $r_0 \gg L$. There is a naked
singularity at $r=0$ and the temperature of the cosmic horizon reads
$T_c\simeq {n \over 4 \pi L} \left({r_o \over L} \right)^{n-2 \over
n}$. Now we have \eqn\nhms{\eqalign{\bigtriangleup M&={1 \over n}
{(4 \pi)^n \over (n-2)^{n-1}} {r_N^{2n-2} \over \wn} T_c^n, \cr
\bigtriangleup S_c&= {1 \over n-1} {(4 \pi)^n \over (n-2)^{n-1}}
{r_N^{2n-2} \over \wn} T_c^{n-1}. }} In this limit, one may
postulate that gravity in spacetime is described by a CFT in
$(n-1)+1$ dimensional spacetime.  Again we find that the entropy in
eq.\nhms\ after extended back to the physical region is always
smaller than that in eq.\ndms. This result shows that this $(n-1)+1$
CFT cannot describe gravity of Schwarzschild-de Sitter spacetime in
$n+1$ dimensions.

Finally, we study the case of the $2+1$ dimensional de Sitter space
(some points appeared previously in \ssv). There is no black hole in
the 2+1 dimensional de Sitter space, however, there exist solutions
also called Schwarzschild-de Sitter with only one horizon and the
metric is given by \eqn\cmet{ds^2=-\left(1-8Gm-{r^2 \over L^2}
\right) dt^2 + \left(1-8Gm-{r^2 \over L^2} \right)^{-1} dr^2 + r^2
d\phi^2, } where $L=1/\sqrt{\Lambda}$ is the size of the pure de
Sitter space with cosmological constant $\Lambda$. There is no black
hole horizon in this spacetime. The size of the cosmic horizon is
\eqn\ccs{r_c=L\sqrt{1-8Gm}, } the temperature of the cosmic horizon
is \eqn\cct{T_c={r_c \over 2 \pi L^2 }={\sqrt{1-8Gm} \over 2 \pi L}
} and the entropy corresponding to the cosmic horizon is
\eqn\centr{S_c={2 \pi r_c \over 4G}={\pi L \sqrt{1-8Gm} \over  2G
}.} The first law of thermodynamics takes the form
\eqn\cfl{d(-m)=T_c d S_c.,} so $-m$ can be viewed as energy. The
spacetime \ccs\ can be thought of as generated by a massive point
particle, and there is a deficit angle. When $m=1/(8G)$, the deficit
angle becomes $2 \pi$ and the space is closed up. For a generic $m$,
we are not supposed to take the system to describe the maximal
entropy state of a given energy, since there is no hole around.
Nevertheless, taking the limit $m=1/(8G)$ we find that the
temperature of the cosmic horizon vanishes and the entropy tends to
zero too, we may postulate that this state is the ground state. The
proper distance from $r=0$ to the cosmic horizon is given by
$\ell=\pi L/2$. The horizon collapses in this limit thus the whole
space collapses to a segment of length $\ell$. One may postulate
that the system is described by a 1+1 dimensional CFT. In fact, in
the whole physical range of $m$, the subtracted energy and the
entropy behave as if there is a CFT \eqn\cms{\eqalign{\bigtriangleup
M&={\pi L \over G} \ell T_c^2, \cr \bigtriangleup S_c&={2 \pi L
\over G}\ell T_c. }} One may be tempted to guess that the whole de
Sitter gravity is dual to a 2d CFT, this is certainly misleading,
since as we said before that a state with a massive particle sitting
at $r=0$ is hardly the generic state of the given energy. However,
the limit $m=1/(8G)$ is special, since the metric collapses to
$ds^2=d \rho^2$, where $\rho=L \sin^{-1} \left({r \over r_c}
\right)$, so one may propose that near this limit there is a 2d CFT
with central charge $c=6L/G$ living on the segment.

To summarize, we take the Nariai black hole as the ground state and
made one conjecture and two observations on the thermodynamical
behavior of Schwarzschild-de Sitter space in three different
regions. We conjecture that near the Nariai limit the system is dual
to a $2d$ CFT living on the finite segment stretched between the
cosmic horizon and the black hole horizon. One observation is that
in the pure de Sitter phase, the system's behavior is similar to the
Hagedorn string in a fixed volume in $2+1$ spacetime and another one
is that the imaginary high temperature phase of quantum gravity in
$n+1$ dimensional dS spacetime is dual to a $(n-1)+1$ dimensional
CFT. Of course it is a long way to go to reach any certain
conclusion about the de Sitter quantum gravity. We hope that the
conjecture and observations made here will be useful in the on-going
search for the foundation of the de Sitter gravity.

\bigskip

Acknowledgments

The work of QGH was supported by a grant from NSFC, a grant
from China Postdoctoral Science Foundation and and a grant from K.
C. Wang Postdoctoral Foundation. The work of KK and ML was supported by a
grant from CAS and a grant from NSFC.

\listrefs
\end